\documentclass[11pt]{article}
\newcommand{\Comment}[1]{{}}
\usepackage{amsfonts,amsthm,amsmath,amssymb,slashed}
\usepackage[textwidth = 430 pt, textheight = 630 pt]{geometry}
\usepackage{color}

\Comment{\usepackage{color}
\definecolor{MyDarkBlue}{rgb}{0.15,0.15,0.45}
\usepackage[linktocpage=true]{hyperref}
\hypersetup{
colorlinks=true,
citecolor=MyDarkBlue,
linkcolor=MyDarkBlue,
urlcolor=MyDarkBlue,
pdfauthor={Cristhiam Lopez-Arcos, Jeff Murugan and  Horatiu Nastase},
pdftitle={Nonrelativistic limit of abelianized ABJM and condensed matter applications},
pdfsubject={hep-th}
}

\usepackage[numbers,sort&compress]{natbib}
\usepackage{hypernat}}
\usepackage{graphicx}
\usepackage{cite}

\newcommand\ignore[1]{}
\def\one{{\,\hbox{1\kern-.8mm l}}}

\def\a{\alpha}

\def\d{\partial}

\def\D{\Delta}

\def\Dslash{D\!\!\!\!/\,\,}

\newcommand{\Cset}{{\,\,{{{^{_{\pmb{\mid}}}}\kern-.45em{\mathrm C}}}}}

\newcommand{\be}{\begin{equation}}
\newcommand{\bea}{\begin{eqnarray}}

\newcommand{\ee}{\end{equation}}
\newcommand{\eea}{\end{eqnarray}}

\newcommand{\nn}{\nonumber}

\begin{document}

\renewcommand{\thefootnote}{\fnsymbol{footnote}}

\makeatletter
\@addtoreset{equation}{section}
\makeatother
\renewcommand{\theequation}{\thesection.\arabic{equation}}

\rightline{}
\rightline{}

\begin{flushright}
QGASLAB-15-07
\end{flushright}

\vspace{10pt}


\begin{center}
{\LARGE \bf{\sc Nonrelativistic limit of the abelianized ABJM model and the ADS/CMT correspondence}}
\end{center} 
 \vspace{1truecm}
\thispagestyle{empty} \centerline{
{\large \bf {\sc Cristhiam Lopez-Arcos${}^{a,}$}}\footnote{E-mail address: \Comment{\href{mailto:crismalo@ift.unesp.br}}{\tt crismalo@ift.unesp.br}},
{\large \bf {\sc Jeff Murugan${}^{a,}$}}\footnote{E-mail address: \Comment{\href{mailto:jeff.murugan@uct.ac.za}}{\tt jeff.murugan@uct.ac.za}}
{\bf{\sc and}}
{\large \bf {\sc Horatiu Nastase${}^{b,}$}}\footnote{E-mail address: \Comment{\href{mailto:nastase@ift.unesp.br}}{\tt nastase@ift.unesp.br}}
                                                          }

\vspace{.5cm}

 
\centerline{{\it ${}^a$
The Laboratory for Quantum Gravity \& Strings, }} \centerline{{\it
Department of Mathematics and Applied Mathematics, }} \centerline{{\it
University of Cape Town, Private Bag, Rondebosch, 7700, South Africa}}

\centerline{{\it ${}^b$ 
Instituto de F\'{i}sica Te\'{o}rica, UNESP-Universidade Estadual Paulista}} \centerline{{\it 
R. Dr. Bento T. Ferraz 271, Bl. II, Sao Paulo 01140-070, SP, Brazil}}

\vspace{1truecm}

\thispagestyle{empty}

\centerline{\sc Abstract}

\vspace{.4truecm}

\begin{center}
\begin{minipage}[c]{380pt}
{\noindent We consider the nonrelativistic limit of the abelian reduction of the massive ABJM model proposed in \cite{Mohammed:2012gi},  obtaining a supersymmetric version of the Jackiw-Pi model. The system exhibits an ${\cal N}=2$ 
Super-Schr\"{o}dinger symmetry with the Jackiw-Pi vortices emerging as BPS solutions. We find that this $(2+1)$-dimensional abelian field theory is dual to a certain (3+1)-dimensional gravity theory that differs somewhat from previously considered abelian condensed matter stand-ins for the ABJM model. We close by commenting on progress in the top-down realization of the AdS/CMT correspondence in a critical string theory. 
}
\end{minipage}
\end{center}

\vspace{.5cm}

\setcounter{page}{0}
\setcounter{tocdepth}{2}

\newpage

\renewcommand{\thefootnote}{\arabic{footnote}}
\setcounter{footnote}{0}

\linespread{1.1}
\parskip 4pt



\section{Introduction}
\ \ \ \ \ 
Due in no small part to its role in the AdS/CFT correspondence \cite{Maldacena:1997re}, $4$-dimensional, $\mathcal{N}=4$ supersymmetric Yang-Mills 
theory (SYM) has provided a remarkable new window into the physics of strongly coupled gauge theories, resulting in a veritable spectrum of insights that range from the phenomenology of the quark-gluon plasma to the structure of scattering amplitudes in quantum field theories. Its $3$-dimensional counterpart, an $\mathcal{N}=6$ superconformal Chern-Simons-matter theory called the ABJM model \cite{Aharony:2008ug}, has furnished an equally impressive laboratory within which to understand planar field theories and promises a powerful toolbox with which to attack various condensed matter systems, again via the gauge/gravity duality.  

This Anti de-Sitter/Condensed-Matter-Theory (AdS/CMT) correspondence is usually defined somewhat phenomenologically, by building gravity duals with the required fields and symmetries to describe relevant physics in (typically) abelian condensed matter models and certainly much of the progress in the field has been made in this {\it bottom-up} approach \cite{Hartnoll:2009sz}. While this progress is certainly remarkable, we felt it unsatisfactory and found the need to ask whether it was possible to realize such a correspondence in a critical string theory, like the type IIA $AdS_4\times \mathbb{CP}^4$. 

This article is a continuation of the {\it top-down} program for the construction of an AdS/CMT correspondence initiated in \cite{Mohammed:2012gi,Mohammed:2012rd}.  There it was demonstrated that a (fully quantum) consistent truncation of (a  massive deformation of) the ABJM model \cite{Gomis:2008vc,Terashima:2008sy}, reduces to a relativistic version of the Landau-Ginzburg model. The latter of course, plays a key role in many condensed matter phenomena, for example, quantum critical phases \cite{Myers:2010pk,Sachdev:2011wg}.
Then, in \cite{Murugan:2013jm}, we extended the truncation to the supersymmetric case, demonstrating its consistency with the supersymmetry 
of the parent (m)ABJM model and its utility in planar condensed matter systems\footnote{For an extension of these methods to $\mathcal{N}=4$ SYM, see \cite{Cardona:2015efa}.}. 

However, while there are certainly condensed matter systems that are {\it effectively} relativistic, most are, in fact, nonrelativistic. 
The primary purpose of this paper is therefore to take the next logical step, and perform a nonrelativistic limit on the abelian reduction found in \cite{Murugan:2013jm}. We find a reduction to the so-called Jackiw-Pi model of\cite{Jackiw:1990mb,Jackiw:1990tz}, with the well studied Jackiw-Pi vortices arising as solutions of the reduced (m)ABJM model. Further, we also find a supersymmetric version of the model defined in \cite{Leblanc:1992wu}, with an ${\cal N}=2$ supersymmetric Schr\"{o}dinger symmetry. This particular reduction allows us to describe regular systems with Schr\"{o}dinger symmetry, via the usual AdS/CFT holography, in terms of a $(d+1)$-dimensional gravity dual. This is to be contrasted with the previously known standard example of holography for Schr\"{o}dinger symmetry, between an unusual dipole theory and a gravity dual in $(d+2)$ dimensions, obtained from the discrete light cone quantization (DLCQ) of usual AdS/CFT dualities
\cite{Maldacena:2008wh,Herzog:2008wg,Adams:2008wt}.
We also compare our model with the abelian nonrelativistic model in \cite{Huijse:2011hp,Huijse:2011ef}, that was used as a stand-in for ABJM to describe compressible Fermi surfaces. 

The paper is organized as follows. In section 2 we consider the nonrelativistic limit of the abelian reduction of the massive ABJM model, and 
explore two choices for the supersymmetry transformation rules. In section 3 we further truncate the model to obtain the 
supersymmetric version of the model of \cite{Leblanc:1992wu}, whose symmetry is coded in an ${\cal N}=2$ Super-Schr\"{o}dinger 
algebra. In section 4 we describe applications to the AdS/CMT correspondence, first by comparing with systems with Schr\"{o}dinger symmetry previously used in this context, and then by comparing with the nonrelativistic model of \cite{Huijse:2011hp}, previously used to understand the physics of compressible Fermi surfaces. We conclude with a discussion in section 5.

\section{A nonrelativistic limit of abelianized ABJM}

Our starting point for this study is the abelian reduction of the mass-deformed ABJM model proposed in \cite{Murugan:2013jm}. The action for the supersymmetric abelian reduction of massive ABJM was found in eq. 4.6 of that paper, and reads\footnote{Note that this version differs from that found in \cite{Murugan:2013jm} by an $i$, a minus sign and 
replacing the $\eta^\dagger$'s with $\bar{\eta}$'s}
\bea
S&=&-\frac{N(N-1)}{2}\int d^3x\Bigg\{\frac{k}{4\pi}\epsilon^{\mu \nu \lambda}\big(a^{(2)}_{\mu}f^{(1)}_{\nu   
  \lambda}+a^{(1)}_{\mu}f^{(2)}_{\nu \lambda}\big)
  +|D_{\mu}\phi_{i}|^{2}+|D_{\mu}\chi_{i}|^{2}\cr
&&+i\sum_{i=1,2}\Big[\bar{\eta}_i (\Dslash +\mu)\eta_i +
\bar{\tilde\eta}_i(\Dslash-\mu)\tilde\eta_i\Big]\cr
&&-\frac{2\pi i}{k}\Big[(|\phi_1|^2+|\chi_1|^2)(\bar{\eta}_2\eta_2+\bar{\tilde\eta}_2\tilde\eta_2)
+(|\phi_2|^2+|\chi_2|^2)(\bar{\eta}_1\eta_1+\bar{\tilde\eta}_1\tilde\eta_1)\Big]\cr
&&+\left(\frac{2\pi}{k}\right)^2\Big[(|\phi_{1}|^{2}+|\chi_{1}|^{2})\big(|\chi_{2}|^{2}-|\phi_{2}|^{2}
  -c^{2}\big)^{2}
  +(|\phi_{2}|^{2}+|\chi_{2}|^{2})\big(|\chi_{1}|^{2}-|\phi_{1}|^{2}-c^{2}\big)^{2}\cr
&+&4|\phi_{1}|^{2}|\phi_{2}|^{2}(|\chi_{1}|^{2}+|\chi_{2}|^{2})+4|\chi_{1}|^{2}|\chi_{2}|^{2}(|\phi_{1}|^{2}+|\phi_{2}|^{2})\Big]\Bigg\}\;,
\label{susyaction}
\eea
where $c^2\equiv k\mu/(2\pi)$, the abelian covariant derivative on the scalars $D_{\mu}\phi_i = \left(\partial_{\mu} - iA_{\mu}^{(i)}\right)\phi_i$, with a similar relation holding for  
fermions, and $\bar{\eta}\equiv \eta^\dagger \gamma^0$. Here $a_\mu^{(1)}, a_\mu^{(2)}$ are gauge fields, $\phi_1, \phi_2$ and 
$\chi_1,\chi_2$ are complex scalars and $\eta_1,\eta_2$ and $\tilde\eta_1,\tilde\eta_2$ are complex 2-component Dirac spinors. This action is invariant under the following set of supersymmetry transformations rules\footnote{Note that we have removed the 1/2 in front of the $\mu $ term in the susy rules 
and multiplied it with a minus sign everywhere, correcting the corresponding result in \cite{Murugan:2013jm}, as is easily checked. We have also 
used $\epsilon^*$ instead of $\epsilon$ in the second terms in $\delta a_\mu^{(i)}$.}
\bea
\delta \phi_1&=& i\bar\epsilon\tilde\eta_{\dot 1},\nonumber\\
\delta \phi_2&=& -i\bar\epsilon\tilde\eta_{\dot 2},\nonumber\\
\delta \chi_{\dot 1}&=& -i\bar\epsilon\eta_{1},\nonumber\\
\delta \chi_{\dot 2}&=& i\bar\epsilon\eta_{2},\nonumber\\
\delta a_\mu^{(1)}&=&\frac{2\pi}{k}\left(\bar \epsilon\gamma_\mu[\phi_2\tilde\eta_{\dot 2}^*-\chi_{\dot 2}\eta_2^*]
+\bar\epsilon^*\gamma[\phi_2^*\tilde \eta_{\dot 2}-\chi^*_{\dot 2}\eta_2]\right),\nonumber\\
\delta a_\mu^{(2)}&=&-\frac{2\pi}{k}\left(\bar \epsilon\gamma_\mu[\phi_1\tilde\eta_{\dot 1}^*-\chi_{\dot 1}\eta_1^*]
+\bar\epsilon^*\gamma^\mu[\phi_1^*\tilde \eta_{\dot 1}-\chi^*_{\dot 1}\eta_1]\right),\\
\delta \eta_1&=&\gamma^\mu D_\mu \chi_{\dot 1}+\frac{2\pi}{k}\epsilon\chi_{\dot 1}(|\phi_2|^2+|\chi_{\dot 2}|^2)-\mu\epsilon\chi_{\dot 1},\nonumber\\
\delta \eta_2&=&-\gamma^\mu \epsilon D_\mu \chi_{\dot 2}-\frac{2\pi}{k}\epsilon\chi_{\dot 2}(|\phi_1|^2+|\chi_{\dot 2}|^2)+\mu\epsilon\chi_{\dot 2},\nonumber\\
\delta \tilde \eta_{\dot 1}&=&-\gamma^\mu \epsilon D_\mu \phi_{ 1}-\frac{2\pi}{k}\epsilon\phi_{1}(|\phi_2|^2+|\chi_{\dot 2}|^2)-\mu\epsilon\phi_{ 1},\nonumber\\
\delta \tilde \eta_{\dot 2}&=&\gamma^\mu \epsilon D_\mu \phi_{ 2}+\frac{2\pi}{k}\epsilon\phi_{2}(|\phi_1|^2+|\chi_{\dot 1}|^2)+\mu\epsilon\phi_{2}.\nonumber
\label{susyrules}
\eea
Since the parameter $\epsilon$ is complex, we have an $SO(2)=U(1)$ R-symmetry, resulting in an ${\cal N}=2$ susy in three dimensions. 

\subsection{A nonrelativistic limit of the action}

The nonrelativistic limit for the nonabelian ${\cal N}=6$ mass-deformed ABJM was first considered in \cite{Lee:2009mm,Nakayama:2009cz}. Here, we will focus on the abelianized ABJM. In order to take the nonrelativistic limit, we first need to restore the factors of $\hbar $ and $c$ by dimensional analysis.\footnote{The 
dimensions of constants and fields in terms of mass $M$, length $L$ and time $T$ are: $[\hbar] = M L^2 T^{-1}$, $[m] = M$, $[c] = L T^{-1}$, 
$[k] = L^{-1} T$, $[Z^{\hat{A}}] = [W^{\dagger \check{A}}] = M^{1/2} L^{1/2} T^{-1/2}$, 
$[\psi_A] = M^{1/2} T^{-1/2}$, $[A_{\mu}] = [\hat{A}_{\mu}] = L^{-1}$, $[A_t] = T^{-1}$.} Writing also ($(\d_0,A_0)=(\d_t,A_t)/c$)
and renaming the nonrelativistic $\mu$ as $m$, the scalar 
part of the Lagrangian becomes
\bea
-\frac{2}{N(N-1)}{\cal L}_{{\rm scal}} &=& -\frac{1}{c^2}(D_t\tilde{\phi}_j)\overline{(D_t\tilde{\phi}_j)} +(D_i\tilde{\phi}_j)\overline{(D_i\tilde{\phi}_j)} 
-\frac{1}{c^2}(D_t\tilde{\chi}_j)\overline{(D_t\tilde{\chi}_j)}\cr
&& +(D_i\tilde{\chi}_j)\overline{(D_i\tilde{\chi}_j)}
 +\frac{m^2c^2}{\hbar^2}\left(|\tilde{\phi}_1|^2 + |\tilde{\phi}_2|^2 + |\tilde{\chi}_1|^2 + |\tilde{\chi}_2|^2\right) \cr
&&- \frac{8\pi}{k}\frac{mc}{\hbar}\frac{1}{\hbar c}\left(|\tilde{\chi}_1|^2|\tilde{\chi}_2|^2 - |\tilde{\phi}_1|^2|\tilde{\phi}_2|^2\right)
+\frac{4\pi^2}{(k\hbar c)^2}\Big[(|\tilde{\chi}_1|^2 + |\tilde{\phi}_1|^2)(|\tilde{\chi}_2|^2 + |\tilde{\phi}_2|^2) \cr
&&\times (|\tilde{\chi}_1|^2 
+ |\tilde{\chi}_2|^2 + |\tilde{\phi}_1|^2 + |\tilde{\phi}_2|^2)\Big],\nn\\
\eea
whereas the pure gauge (Chern-Simons) part of the Lagrangian is topological, so it is unchanged, up to the fact that now there is a 
relative $c$ between the spatial and temporal parts of the action,
\bea
S_{CS}^{NR}=-\frac{N(N-1)}{2}\int d^3x\Bigg\{\frac{k\hbar}{4\pi}\epsilon^{\mu \nu\lambda}\big(A^{(2)}_{\mu}F^{(1)}_{\nu\lambda} 
+ A^{(1)}_{\mu}F^{(2)}_{\nu \lambda}\big)\Bigg\}.
\label{nrcs}
\eea
Note that the overall factor of $c$ is cancelled by the re-definition of $A_0$ in the nonrelativistic, $c\rightarrow \infty$ limit, which also eliminates the sextic terms in the scalar potential.
For the remaining terms, we must replace the fields with their nonrelativistic versions. In principle, a complex scalar field $\tilde{\phi}$ would be written as
\be
\tilde \phi=\frac{\hbar}{\sqrt{2m}}\left[\phi e^{-i\frac{mc^2}{\hbar}t}+\hat\phi^* e^{+i\frac{mc^2}{\hbar}t}\right]\;,
\ee
with $\phi$ and $\hat\phi^*$ complex fields representing particles and anti-particles respectively, separately conserved. However, we will be working in the 
zero antiparticle sector, where we drop the second term so that 
\bea
(\tilde{\phi},\tilde{\chi})\longrightarrow \left(\frac{\hbar}{\sqrt{2m}}\phi(x,t)e^{-imc^2t/\hbar},\frac{\hbar}{\sqrt{2m}}\chi(x,t)e^{-imc^2t/\hbar}\right).
\eea
From the kinetic (time derivative) term, the purely scalar part is 
\bea
-\frac{\hbar^2}{2mc^2}\left(\d_t\phi\d_t\bar{\phi} + \frac{i2mc^2}{\hbar}\bar{\phi}\d_t\phi +\frac{m^2c^4}{\hbar^2}|\phi|^2\right).
\eea
Of the three terms present, the first does not survive the nonrelativistic limit, and the last one cancels the mass term. On the other hand, the terms containing the interaction with $A_t$ are
\bea
-\frac{\hbar^2}{2mc^2}\left(\frac{2mc^2}{\hbar}\phi A_t\bar{\phi} +iA_t\bar{\phi}\d_t\phi -iA_t\phi\d_t\bar{\phi} +A_t^2|\phi|^2\right).
\eea
Here only the first term survives so that the kinetic term for the scalar contributes in total
\bea
-\bar{\phi}i\hbar D_t\phi.
\eea
Putting everything together, gives the nonrelativistic scalar action
\bea
S^{NR}_{scal}&=&-\frac{N(N-1)}{2}\int dx^3\Bigg\{-\bar{\phi}_{i}\left(i\hbar D_t + \frac{\hbar^2}{2m}D^2_i \right)\phi_i 
- \bar{\chi}_{i}\left(i\hbar D_t + \frac{\hbar^2}{2m}D^2_i \right)\chi_i\nn\\
&&\hspace{3.5cm} + \frac{2\pi\hbar^2}{mk}\left(|\phi_1|^2|\phi_2|^2 - |\chi_1|^2|\chi_2|^2\right)\Bigg\}.
\label{nrscal}
\eea
Note that for $k>0$, the $\chi$'s have negative potential after the nonrelativistic limit. The positive sextic terms, which would have regulated this dependence and restored the positivity of the potential, do not survive the limit. 
Evidently then, for $k>0$ we seem to need  (at least one of the) $\chi_i=0$ for the consistency of the theory, while $k<0$ requires (at least one of the)$\phi_{i}=0$.

Let us go now to the fermionic sector of the theory where, with the factors of $\hbar$ and $c$ included, the Lagrangian reads
\bea
-\frac{2}{N(N-1)}{\cal L}_{{\rm fer}} &=& i\sum_{i=1,2}\Big[\bar{\eta}_i (\Dslash +\frac{mc}{\hbar})\eta_i +
\bar{\tilde\eta}_i(\Dslash -\frac{mc}{\hbar})\tilde\eta_i\Big]\nn\\
&&-\frac{2\pi i}{k\hbar c}\Big[(|\phi_1|^2+|\chi_1|^2)(\bar{\eta}_2\eta_2+\bar{\tilde\eta}_2\tilde\eta_2)
+(|\phi_2|^2+|\chi_2|^2)(\bar{\eta}_1\eta_1+\bar{\tilde\eta}_1\tilde\eta_1)\Big].\nn\\
\eea
Again we should really write the nonrelativistic version of the fermion fields as 
\be
\tilde \eta=\sqrt{\hbar c} \left[\psi e^{-i\frac{mc^2}{\hbar}t}+\sigma_2\hat\psi^* e^{+i\frac{mc^2}{\hbar}t}\right]\;,
\ee
where $\psi$ and $\hat\psi$ are complex fields corresponding to particles and anti-particles respectively. In the zero antiparticle sector then,
\bea
\eta_i = \sqrt{\hbar c}\psi_i(x,t)e^{-i\frac{mc^2}{\hbar}t}.
\label{nrfer}
\eea
For the three-dimensional gamma matrices we choose the representation in which
\bea
\gamma^0 = i\tau^3\, ,\,\gamma^1 = \tau^1 \, ,\,\gamma^2 = -\tau^2,
\eea
so that 
\bea
i\bar{\eta}\Dslash\eta=i\eta^\dagger\gamma^0\left(\frac{1}{c}\gamma^0 D_t+\gamma^iD_i\right)\eta
=\hbar c\psi^\dagger\left(-\frac{1}{c}D_t +i\frac{mc}{\hbar}+\gamma^0\gamma^iD_i\right)\psi
\eea
and $\gamma^0\gamma^1=-\tau^2$, $\gamma^0\gamma^2=-\tau^1$. 
Consequently,  
\be
i\gamma^0\gamma^iD_i=i\gamma^0\gamma^1 D_1+i\gamma^0\gamma^2D_2=\begin{pmatrix}0&-D_+\\ D_-& 0\end{pmatrix}\;,
\ee
where $D_\pm\equiv D_1\pm iD_2$. In the nonrelavistic limit only half of the fermion components remain dynamical. For brevity, we will analyze the term with positive mass with an analogous analysis holding for the negative mass case. 
Substituting \eqref{nrfer} into the kinetic term in $-2{\cal L}/N(N-1)$, gives it the form 
\bea
\begin{pmatrix}
\psi^{\dagger}_{i,1} & \psi^{\dagger}_{i,2}
\end{pmatrix}
\begin{pmatrix}
  -i\hbar D_t - 2mc^2 & -\hbar cD_+ \\ \hbar cD_- & -i\hbar D_t
\end{pmatrix}
\begin{pmatrix}
  \psi_{i,1} \\ \psi_{i,2}
\end{pmatrix}.\label{fermikin}
\eea
The ensuing equations of motion
\bea
i\hbar D_t\psi_{i,1} + 2mc^2\psi_{i,1} + \hbar cD_+\psi_{i,2} &=& 0\cr
\hbar cD_-\psi_{i,1} - i\hbar D_t\psi_{i,2}&=& 0,\label{psieqs}
\eea
substantiate our claim above that only half of the fermion components are dynamical, since we can 
solve $\psi_{i,1}$ in terms of $\psi_{i,2}$ by taking only the leading order contribution:
\bea
\psi_{i,1} = -\frac{\hbar}{2mc}D_+\psi_{i,2} - i\frac{\hbar}{2mc^2}D_t\psi_{i,1}.
\eea
The equation of motion for $\psi_{i,2}$ is the Pauli equation for nonrelativistic fermions:
\bea
-i\hbar D_t\psi_{i,2} - \frac{\hbar^2}{2m}D_-D_+\psi_{i,2} + \mathcal{O}\left(\frac{1}{c}\right) = 0,
\eea
rewritten, using $D_-D_+=D_1D_1+D_2D_2+i[D_1,D_2]=D_jD_j+F_{12}$, as
\bea
i\hbar D_t\psi_{i,2} = -\frac{\hbar^2}{2m}D_-D_+\psi_{i,2} = -\frac{\hbar^2}{2m}\left(D_jD_j + F_{12}\right)\psi_{i,2}.\label{psi2eom}
\eea
This equation should be obtained from the action for the fermions.
In the action, the terms in $-2{\cal L}/N(N-1)$ are rewritten as
\bea
&&\psi_{i,1}^\dagger(-i\hbar D_t)\psi_{i,1}+\psi_{i2}^\dagger (-i\hbar D_t)\psi_{i,2}+\hbar c (\psi_{i,2}^\dagger D_-\psi_{i,1}-\psi_{i,1}^\dagger
D_+\psi_{i,2})-2mc^2\psi_{i,1}^\dagger \psi_{i,1}\cr
&=&\psi_{i,2}^\dagger(-i\hbar D_t-\frac{\hbar^2}{2m}D_-D_+)\psi_{i,2}\cr
&=&-i\hbar \psi_{i,2}^\dagger D_t\psi_{i,2}+\frac{\hbar^2}{2m}|D_j \psi_{i,2}|^2-F_{12}\psi^\dagger_{i,2}\psi_{i,2}\;,
\eea
where in the second line we have substituted $\D_+\psi_{i,2}$ from the first eq. in (\ref{psieqs}) and $\psi_{i,1}$ in $D_i\psi_{i,1}$ from the same equation, 
and dropped terms that vanish in the $c\rightarrow \infty$ limit, and in the third we have used $D_-D_+=D_jD_j+F_{12}$ and partially integrated one 
$D_j$. The same analysis carried out for the $\tilde{\eta}$ fermions leads to the kinetic terms
\be
\begin{pmatrix}\tilde{\psi}^\dagger_{i,1}&\tilde{\psi}^\dagger_{i,2}\end{pmatrix}\begin{pmatrix}-i\hbar D_t &-\hbar c D_+\\
\hbar c D_-&-i\hbar D_t -2mc^2\end{pmatrix}\begin{pmatrix}\psi_{i,1}\\ \psi_{i,2}\end{pmatrix}\;,
\ee
which give the similar result,
\bea
\tilde{\psi}_{i,2} &=& \frac{\hbar}{2mc}D_-\tilde{\psi}_{i,1},\nn\\
i\hbar D_t\tilde{\psi}_{i,1} &=& -\frac{\hbar^2}{2m}D_+D_-\tilde{\psi}_{i,1} = -\frac{\hbar^2}{2m}\left(D_jD_j - F_{12}\right)\tilde{\psi}_{i,1}.
\eea
All in all, in the nonrelativistic limit, the fermions are written as
\bea
\tilde{\eta}_i\longrightarrow 
\sqrt{\hbar c}e^{-i\frac{mc^2}{\hbar}t}\begin{pmatrix}
  \tilde{\psi}_{i1} \\ \frac{\hbar}{2mc}D_-\tilde{\psi}_{i1}
\end{pmatrix}\,\,\, ,\,\,\,
{\eta}_i\longrightarrow 
\sqrt{\hbar c}e^{-i\frac{mc^2}{\hbar}t}\begin{pmatrix}
  -\frac{\hbar}{2mc}D_+\psi_{i2} \\ \psi_{i2}
\end{pmatrix}.
\eea
For the ``Yukawa" terms (scalar coupling to fermions) on the other hand, we have the replacement
\bea
-i\bar \eta_i\eta_i&=&\eta_i^\dagger \tau_3\eta_i\rightarrow -\psi_i^\dagger \psi_i\cr
-i\tilde{\bar \eta}_i\tilde\eta_i&=&\tilde\eta_i^\dagger\tau_3\tilde\eta_i\rightarrow +\tilde\psi_i^\dagger\tilde\psi_i.
\eea
In what follows, we will drop the 1 and 2 indices on the fermions with the understanding that only one component survives the nonrelativistic limit.

\noindent
After replacing the fields in the action with their nonrelativistic avatars, the fermionic part of the action takes the form
\bea
S^{NR}_{fer}&=&-\frac{N(N-1)}{2}\int dx^3\Bigg\{\sum_{j=1,2}\Big[-\psi^\dagger_j\left(i\hbar D_t + \frac{1}{2m}\left(D_i^2 + F_{12}^{(j)}\right)\right)\psi_j\nn\\ 
&&\hspace{3.5cm}- \tilde{\psi}^\dagger_j\left(i\hbar D_t + \frac{1}{2m}\left(D_i^2 - F_{12}^{(j)}\right)\right)\tilde{\psi}_j\Big]\nn\\
&&\hspace{3.5cm} - \frac{\pi\hbar^2}{km}\Big[(|\phi_1|^2+|\chi_1|^2)(\psi^\dagger_2\psi_2-\tilde{\psi}^\dagger_2\tilde{\psi}_2)\nn\\
&&\hspace{3.5cm}+(|\phi_2|^2 + |\chi_2|^2)(\psi^\dagger_1\psi_1 - \tilde{\psi}^\dagger_1\tilde{\psi}_1)\Big].
\Bigg\}.
\label{nrferm}
\eea
Together, the equations \eqref{nrcs}, \eqref{nrscal}, and \eqref{nrferm} furnish the full nonrelativistic abelianized massive ABJM action.

\subsection{Nonrelativistic limit of the full susy rules}

As a check, we now attempt to take the same nonrelativistic limit at the level of the supersymmetry transformation rules \eqref{susyrules}. Reintroducing $\hbar$ and $c$, by replacing $\mu$ with $mc/\hbar$, $k$ with $kc$ and $D_\mu$ with $D_t/c+D_i$, we get 
\bea
\delta \phi_1&=& i\bar\epsilon\tilde\eta_{\dot 1},\nn\\
\delta \phi_2&=& -i\bar\epsilon\tilde\eta_{\dot 2},\nn\\
\delta \chi_{\dot 1}&=& -i\bar\epsilon\eta_{1},\nn\\
\delta \chi_{\dot 2}&=& i\bar\epsilon\eta_{2},\nn\\
\delta A_\mu^{(1)}&=&\frac{2\pi}{kc}\left(\bar \epsilon\gamma_\mu[\phi_2\tilde\eta_{\dot 2}^*-\chi_{\dot 2}\eta_2^*]
+\bar\epsilon^*\gamma[\phi_2^*\tilde \eta_{\dot 2}-\chi^*_{\dot 2}\eta_2]\right),\nn\\
\delta A_\mu^{(2)}&=&-\frac{2\pi}{kc} \left(\bar \epsilon\gamma_\mu[\phi_1\tilde\eta_{\dot 1}^*-\chi_{\dot 1}\eta_1^*]
+\bar\epsilon^*\gamma^\mu[\phi_1^*\tilde \eta_{\dot 1}-\chi^*_{\dot 1}\eta_1]\right),\\
\delta \eta_1&=&\gamma^\mu \epsilon D_\mu \chi_{\dot 1}+\frac{2\pi}{kc}\epsilon\chi_{\dot 1}(|\phi_2|^2+|\chi_{\dot 2}|^2)-\frac{mc}{\hbar}
\epsilon\chi_{\dot 1},\nn\\
\delta \eta_2&=&-\gamma^\mu \epsilon D_\mu \chi_{\dot 2}-\frac{2\pi}{kc}\epsilon\chi_{\dot 2}(|\phi_1|^2+|\chi_{\dot 2}|^2)+\frac{mc}{\hbar}
\epsilon\chi_{\dot 2},\nn\\
\delta \tilde \eta_{\dot 1}&=&-\gamma^\mu \epsilon D_\mu \phi_{ 1}-\frac{2\pi}{kc}\epsilon\phi_{1}(|\phi_2|^2+|\chi_{\dot 2}|^2)-\frac{mc}{\hbar}
\epsilon\phi_{ 1},\nn\\
\delta \tilde \eta_{\dot 2}&=&\gamma^\mu \epsilon D_\mu \phi_{ 2}+\frac{2\pi}{kc}\epsilon\phi_{2}(|\phi_1|^2+|\chi_{\dot 1}|^2)+\frac{mc}{\hbar}
\epsilon\phi_{2},\nn
\eea
where here $\delta A_\mu$ is understood as $(\frac{1}{c}\delta A_0, \delta A_i$), and $\gamma^\mu D_\mu=\frac{1}{c} \gamma^0D_0+\gamma^i D_i$. 
Since in the nonrelativistic limit, one of the components of the fermions goes to zero, the same has to happen in the susy transformation rules: the 
variation of the component that goes to zero should also go to zero, and only the variation of the other component should be finite.

\noindent
Note that $\epsilon$ is a complex 2-component spinor. Since a minimal spinor in 3 dimensions is Majorana, with only one independent complex (or two real)
component(s), these susy rules correspond to ${\cal N}=2$ supersymmetry.
These components, which we denote by $\epsilon_1$  (upper) and $\epsilon_2$ (lower) respectively, are to be understood as the independent supersymmetries in the nonrelativistic limit. We first consider the transformation rule for the scalar $\phi_1$,
\bea
\delta\phi_1 = -\sqrt{\frac{2mc}{\hbar}}\left(\epsilon_1^*\tilde{\psi}_{1,1} - \epsilon_2^*\frac{\hbar}{2mc}D_-\tilde{\psi}_{1,1}\right).\label{scalartransf}
\eea
Since $c\rightarrow\infty$, both terms are singular in the nonrelativistic limit. In order to circumvent this behaviour, we need to rescale the supersymmetry parameters. This rescaling is not unique. One possible choice for a rescaling of the susy parameters is
\be
\epsilon_i\rightarrow \sqrt{\frac{\hbar }{2mc}}\epsilon_i;\;\;\;\; i=1,2.
\ee
In that case, for the variations of the scalars we obtain
\bea
\delta \phi_1&=&-\epsilon_1^*\tilde \psi_{1,1}\nn\\
\delta\phi_2&=&+\epsilon_1^*\tilde \psi_{2,1}\\
\delta\chi_1&=&-\epsilon_2^*\psi_{1,2}\nn\\
\delta\chi_2&=&+\epsilon_2^*\psi_{2,2},\nn
\eea
while for the fermion variations,
\bea
\delta\begin{pmatrix}-\frac{\hbar }{2mc}D_+\psi_{1,2}\\ \psi_{1,2}\end{pmatrix}&\simeq& (\tau_3\epsilon-\epsilon)\chi_1\nn\\
\delta\begin{pmatrix}-\frac{\hbar }{2mc}D_+\psi_{2,2}\\ \psi_{2,2}\end{pmatrix}&\simeq&- (\tau_3\epsilon-\epsilon)\chi_2\nn\\
\delta\begin{pmatrix}\tilde\psi_{1,1}\\ \frac{\hbar}{2mc}D_-\tilde\psi_{1,1}\end{pmatrix}&\simeq& -(\tau_3\epsilon+\epsilon)\phi_1\\
\delta\begin{pmatrix}\tilde\psi_{2,1}\\ \frac{\hbar}{2mc}D_-\tilde\psi_{2,1}\end{pmatrix}&\simeq& +(\tau_3\epsilon+\epsilon)\phi_1\nn
\eea
Clearly in the nonrelativistic limit the same half of the components vanish on the left hand side and on the right hand side, 
as it should be. The other half gives
\bea
\delta\psi_{1,2}&=&\epsilon_2\chi_1\nn\\
\delta\psi_{2,2}&=&-\epsilon_2\chi_2\nn\\
\delta\tilde\psi_{1,1}&=&-\epsilon_1\phi_1\\
\delta\tilde \psi_{2,1}&=&+\epsilon_1\phi_2.\nn
\eea
Finally, the variations of the gauge fields are expressed as
\bea
\delta A_0^{(1)}&=&\frac{2\pi \hbar^2}{2mk}(\epsilon_1^*\tilde \psi_{2,1}\phi_2^*-\epsilon_2^*\psi_{2,2}\chi_2^*)+c.c,\nn\\
\delta A_i^{(1)}&=&0,\nn\\
\delta A_0^{(2)}&=&-\frac{2\pi \hbar^2}{2mk}(\epsilon_1^*\tilde \psi_{1,1}\phi_2^*-\epsilon_2^*\psi_{2,2}\chi_1^*)+c.c,\\
\delta A_i^{(2)}&=&0.\nn
\eea
Certainly then, the reduction passes this check at the level of the supersymmetry transformations. However, as we have already seen, at the level of the action, when $k>0$ we have 
a negative potential for $\chi$ and for $k<0$, a negative potential for $\phi$, signalling a possible instability. 

\noindent
At this juncture, it is worth noting that the two susies act on $(\phi,\tilde\psi, A_\mu^{(1,2)})$ and $(\chi,\psi, A_\mu^{(1,2)})$ respectively, with the action on the $A_\mu^{(1,2)}$ being specifically a nonlinear one only. At the level of the linearized action, the two supersymmetries evidently act on different fields. Therefore in some sense this 
corresponds to two different sets of ${\cal N}=1$ invariant fields put together.

\subsection{Truncating the susy rules and the action}

If, however, we would like to keep {\it both} terms in the transformation (\ref{scalartransf}) finite, another rescaing that is afforded to us is 
\bea
(\epsilon_1,\epsilon_2)\longrightarrow \left(\sqrt{\frac{\hbar}{2mc}}\epsilon_1,\sqrt{\frac{c}{2m\hbar}}\epsilon_2\right).
\eea
Then the transformation rule for $\phi_1$ takes the form
\bea
\delta\phi_1 = -\epsilon_1^*\tilde{\psi}_{1,1} + \frac{1}{2m}\epsilon_2^*D_-\tilde{\psi}_{1,1}.
\eea
The first term on the right hand side is called {\it kinematical} supersymmetry transformation $\delta_K\phi_1$, and  the second a 
{\it dynamical} one which we denote $\delta_D\phi_1$, with similar rules holding for $\phi_2$. However, a problem appears when 
we consider $\chi_{1,2}$. For example, the transformation rule for $\chi_1$, 
\bea
\delta\chi_1 = -\frac{c}{\hbar}\epsilon_2^*\psi_{1,2}-\frac{1}{2m}\frac{\hbar}{c}\epsilon_1^*D_-\psi_{1,2}.
\eea
implies that, in order to have supersymmetry with {\em both} kinematical and dynamical terms 
in the nonrelativistic abelian case, we are forced to truncate the model by setting
$\chi_i= \psi_i = 0$. Since in this case we will be left with only one set of $\psi$s, we will remove the tilde for simplicity from now on.
With this truncation and rescaling of supersymmetry parameters, the truncated action becomes
\bea
S^{NR}&=&-\frac{N(N-1)}{2}\int d^3x\Bigg\{\frac{k\hbar}{4\pi}\epsilon^{\mu \nu\lambda}\big(A^{(2)}_{\mu}F^{(1)}_{\nu\lambda}
 + A^{(1)}_{\mu}F^{(2)}_{\nu \lambda}\big) - \bar{\phi}_{i}\left(i\hbar D_t + \frac{\hbar^2}{2m}D^2_j \right)\phi_i\nn\\&&\hspace{3.5cm}
 - \sum_{j=1,2}\Big[\psi^\dagger_j\left(i\hbar D_t + \frac{1}{2m}\left(D_i^2 - F_{12}^{(j)}\right)\right)\psi_j\Big]\nn\\
&&\hspace{3.5cm} + \frac{\pi\hbar^2}{km}\Big[(|\phi_1|^2)(\psi^\dagger_2\psi_2)+(|\phi_2|^2)(\psi^\dagger_1\psi_1)\Big] 
+ \frac{2\pi\hbar^2}{mk}\left(|\phi_1|^2|\phi_2|^2 \right)\Bigg\}\;,\cr
&&
\label{nrabjm}
\eea
with the supersymmetry transformation rules,
\bea
\delta\phi_1 &=& -\epsilon_1^*\psi_{1,1} + \frac{1}{2m}\epsilon_2^*D_-\psi_{1,1},\nn\\
\delta\phi_2 &=& \epsilon_1^*\psi_{2,1} - \frac{1}{2m}\epsilon_2^*D_-\psi_{2,1},\nn\\
\delta A_t^{(1)} &=&+\frac{\pi\hbar}{mk}(\epsilon_1^*\phi_2\psi_{2,1}^* + \epsilon_1\phi_2^*\psi_{2,1}) 
+ \frac{2\pi\hbar}{(2m)^2k}(\epsilon_2^*\phi_2D_+\psi_{2,1}^* + \epsilon_2\phi_2^*D_-\psi_{2,1}),\nn\\
\delta A_1^{(1)} &=& -\frac{i\pi\hbar}{mk}(\epsilon_2^*\phi_2\psi_{2,1}^* + \epsilon_2\phi_2^*\psi_{2,1}),\nn\\
\delta A_2^{(1)} &=& -\frac{\pi\hbar}{mk}(\epsilon_2^*\phi_2\psi_{2,1}^* + \epsilon_2\phi_2^*\psi_{2,1}),\\
\delta A_t^{(2)} &=&-\frac{\pi\hbar}{mk}(\epsilon_1^*\phi_1\psi_{1,1}^* + \epsilon_1\phi_1^*\psi_{1,1}) 
- \frac{2\pi\hbar}{(2m)^2k}(\epsilon_2^*\phi_1D_+\psi_{1,1}^* + \epsilon_1\phi_1^*D_-\psi_{1,1}),\nn\\
\delta A_1^{(2)} &=& \frac{i\pi\hbar}{mk}(\epsilon_2^*\phi_1\psi_{1,1}^* + \epsilon_2\phi_1^*\psi_{,11}),\nn\\
\delta A_2^{(1)} &=& \frac{\pi\hbar}{mk}(\epsilon_2^*\phi_1\psi_{1,1}^* + \epsilon_2\phi_1^*\psi_{1,1}),\nn\\
\delta\psi_{1,1} &=& \frac{1}{2m}\epsilon_2D_-\phi_1 - \epsilon_1\phi_1,\nn\\
\delta\psi_{2,1} &=& -\frac{1}{2m}\epsilon_2D_-\phi_2 +\epsilon_1\phi_2,\nn
\eea
We also note the intermediate result for the fermion variation
\be
\delta\begin{pmatrix}\psi_{1,1}\\ \frac{\hbar ^2}{2mc}D_-\psi_{1,1}\end{pmatrix}=\sqrt{\frac{\hbar}{2mc}}\left[
-\frac{1}{c}\gamma^0 \epsilon D_0\phi_1-\gamma^i \epsilon D_i\phi_1-\frac{mc}{\hbar }\phi_1(\tau_3\epsilon+\epsilon)-\frac{\pi \hbar^2}{mkc}
|\phi_2|^2\phi_1\right].
\ee
with a similar one for $\psi_{2,1}$, where 
\be
\epsilon=\begin{pmatrix}\sqrt{\frac{\hbar}{2mc}}\epsilon_1\\ \sqrt{\frac{c}{2m\hbar}}\epsilon_2\end{pmatrix}.
\ee
Then we see that the first and last terms vanish as $c\rightarrow\infty$, whereas the remaining $(\tau_3\epsilon+\epsilon)$ and the $\gamma^i\epsilon D_i$
terms correctly vanish for the lower component only, as it should be, by comparison with the left hand side.

\section{The (supersymmetric) Jackiw-Pi model}

\subsection{The Jackiw-Pi model and its vortex solutions}
In a remarkable series of papers in the early 1990's, beginning with \cite{Jackiw:1990mb}, Jackiw and Pi undertook a systematic study of 
the classical and quantum properties of the gauged nonlinear Schr\"odinger equation\footnote{Our notation in this subsection only will 
match the original literature instead of the rest of this article.}
\begin{eqnarray}
  iD_{t}\psi = -\frac{1}{2}D_{i}^{2}\psi - g\bar{\psi}\psi\psi,
\end{eqnarray} 
for a charged scalar, $\psi$ coupled to an abelian Chern-Simons gauge field whose dynamics is governed by 
\begin{eqnarray}
  \frac{1}{2}\epsilon^{\mu\nu\lambda}F_{\nu\lambda} = \frac{1}{\kappa}j^{\mu},
\end{eqnarray}
and with Chern-Simons coupling (or topological mass) $\kappa$. These equations derive from the Lagrangian density
\begin{eqnarray}
  \mathcal{L} = \frac{\kappa}{4}\epsilon^{\mu\nu\lambda}A_{\mu}F_{\nu\lambda}
  +i\bar{\psi}D_{t}\psi - \frac{1}{2}|D_{i}\psi|^{2}+\frac{g}{2}(\bar{\psi}\psi)^{2},
  \label{JP-lagrangian}
\end{eqnarray}
which defines the so-called Jackiw-Pi model, which has seen enormous development over the past twenty five years as much for its 
pedagogical value in teaching us about four dimensional field theories as for the role that it plays in planar condensed matter 
systems like the quantum Hall effect. For the specific value of the scalar coupling $g = 1/|\kappa|$, the theory takes on a 
``self-dual" structure with the (static) equations of motion descending to the first order set of Bogomolnyi-like equations
\begin{eqnarray}
  D_{i}\psi &=& i\epsilon_{ij}D_{j}\psi,\nonumber\\
  \epsilon_{ij}\,\partial_{i}A_{j} &=& -\frac{1}{\kappa}\bar{\psi}\psi,\nonumber
\end{eqnarray}
supplemented by the Chern-Simons Gauss law constraint that any solution carrying charge $Q$ also possess a 
magnetic flux $\Phi = - Q/\kappa$. These equations are solved exactly by taking the ansatz $\psi = \sqrt{\rho}\,e^{i\omega}$, 
and writing the first order system as a Liouville equation 
\begin{eqnarray}
  \nabla^{2}\ln\rho = -\frac{2}{\kappa}\rho,
\end{eqnarray}  
for the square modulus of the complex scalar. This equation admits a general solution in terms of a holomorphic 
function $f(z)$ of the complex coordinate $z=r\,e^{i\theta}$ on the plane as
\begin{eqnarray}
  \rho(r) = \frac{4\kappa|f'(z)|^{2}}{(1+|f(z)|^{2})^{2}}.
\end{eqnarray}
As a specific example, the choice $f(z) = c_{0}z^{-n}$ yields the axially symmetric solution
\begin{eqnarray}
  \psi(r) = \frac{2\sqrt{\kappa} n}{r}\left(\left(\frac{r_{0}}{r}\right)^{n} + 
  \left(\frac{r}{r_{0}}\right)^{n}\right)^{-1}e^{i(1-n)\theta},
  \label{JP-vortices}
\end{eqnarray}
where the integration constants $r_{0}$ and $n$ are interpreted, respectively, as a scale parameter and a topological charge. 
This corresponds to an $n$-vortex solution located at the origin, the so-called Jackiw-Pi vortex.

\subsection{Nonrelativistic vortices in ABJM}
Returning now to the problem at hand\footnote{And reverting again to our usual notation.}, we consider the bosonic 
part of the nonrelativistic action (\ref{nrabjm}),
\bea
S^{NR}_{bos}&=&-\frac{N(N-1)}{2}\int d^3x\Bigg\{\frac{k\hbar}{4\pi}\epsilon^{\mu \nu\lambda}\big(A^{(2)}_{\mu}F^{(1)}_{\nu\lambda} 
+ A^{(1)}_{\mu}F^{(2)}_{\nu \lambda}\big) +\bar{\phi}_{i}\left(i\hbar D_t + \frac{\hbar^2}{2m}D^2_j \right)\phi_i\nn\\
&&\hspace{3.5cm} + \frac{2\pi\hbar^2}{mk}\left(|\phi_1|^2|\phi_2|^2 \right)\Bigg\},
\eea
and take as an ansatz for a further reduction of the model,
\bea
A_{\mu}^{(1)} &=& A_{\mu}^{(2)} = A_{\mu},\nn\\
\phi_1 &=& \phi_2 = \phi.
\eea
Substituting this into the action leads to
\bea
S_{JP}&=&-N(N-1)\int d^3x\Bigg\{\frac{k\hbar}{4\pi}\epsilon^{\mu \nu\lambda}A_{\mu}F_{\nu\lambda} - \bar{\phi}\left(i\hbar D_t 
+ \frac{\hbar^2}{2m}D^2_j\right)\phi + \frac{\pi\hbar^2}{mk}\left(\phi\bar{\phi})^2 \right)\Bigg\},\nn\\
\eea
which, up to an overall factor of $N^{2}-N$, is just the action for the Jackiw-Pi model (\ref{JP-lagrangian}) encountered a
bove \cite{Jackiw:1990mb,Jackiw:1990tz} and, as such, clearly admits all of the latter's solitonic solutions including, 
the self-dual $n$-vortex Jackiw-Pi vortices (\ref{JP-vortices}). We now show how to understand these vortices in the present context.\\

\noindent
The authors of \cite{Kawai:2009rc} found a class of vortex solutions in the nonrelativistic limit of the massive ABJM action first 
considered by \cite{Lee:2009mm}. These are, in fact, nothing but the Jackiw-Pi vortex, embedded in the ABJM model via the 
abelianization ansatz in \cite{Murugan:2013jm}. Indeed, their solution (eq. (68) of \cite{Kawai:2009rc}) is in our 
notation
\bea
Q^\a(x)&=&\phi(x) G^\a,\nn\\
R^\a(x)&=&\chi(x) G^\a,\nn\\
A_\mu(x)&=&a_\mu(x) G^\a G^\dagger_\a,\\
\hat A_\mu(x)&=&a_\mu(x) G^\dagger_\a G^\a,\nn
\eea
which is just the abelian reduction ansatz in \cite{Murugan:2013jm}, together with the restriction
$\phi_1=\phi_2=\phi$, $\chi_1=\chi_2=\chi$ and $a_\mu^{(1)}=a_\mu^{(2)}=a_\mu$. This is, of course, the same condition 
we administered for comparison with the Jackiw-Pi Lagrangian
(for $\chi=0$). In this case, the BPS equations reduce to
\be
(D_1-iD_2)\phi(x)=0,\;\;\;
(D_1+iD_2)\chi(x)=0\;,
\ee
giving two different types of solutions (referred to as ``BPS I" and ``BPS II" in \cite{Kawai:2009rc}) depending on whether 
either $\chi$ or $\phi$ is turned off. The BPS I vortex solutions are then found from the ansatz $\chi=0$ with
\be
\phi(x)=e^{i\theta(x)}\rho(x)^{1/2}\;,
\ee
which leads to 
\bea
\rho(x)&=&\frac{k}{2\pi}\nabla^2\ln (1+|f(z)|^2)\cr
\theta(x)&=&-(n-1)\arctan(x_2/x_1)\;,
\eea
where $f(z)$ is a holomorphic function of $z=x_1+ix_2$. The BPS II solutions on the other hand, are given by $\phi=0$ and
\be
\chi(x)=e^{i\theta(x)}\rho(x)^{1/2}\;,
\ee
and
\bea
\rho(x)&=&-\frac{k}{2\pi}\nabla^2\ln (1+|f(z)|^2)\cr
\theta(x)&=&(n-1)\arctan(x_2/x_1)\;.
\eea
It was demonstrated in \cite{Kawai:2009rc} that these vortex solutions are indeed BPS, i.e.
they break one conformal, one dynamical and five kinematical supersymmetries, {\it i.e.} exactly half of the 2 conformal, 
2 dynamical and 10 kinematical supersymmetries of the full theory. As we will see shortly, this remains true in our case. 
Since in our case (after the abelian reduction), we have
only ${\cal N}=2$ supersymmetry, i.e. 4 supercharges (2 dynamical and 2 kinematical) and 2 conformal supercharges, the vortices will break 
half of those, i.e. one conformal, one dynamical and one kinematical supersymmetries. 

\subsection{BPS Chern-Simons matter vortices and Jackiw-Pi vortices}

An ${\cal N}=2$  supersymmetric version of the Jackiw-Pi model was considered by Leblanc et al. \cite{Leblanc:1992wu}. Recently, 
in \cite{Tong:2015xaa}, the quantum Hall effect for this gauge theory was studied. The model possesses several remakable properties 
that will be explored in the next subsection. For now, we show that the same theory can be obtained from the ABJM model in our 
nonrelativistic abelian reduction, only with different couplings. Indeed, with the reduction ansatz 
$\phi_1=\phi_2\equiv\phi$, $\psi_1=-\psi_2\equiv \psi$, $A_\mu^{(1)}=A_\mu^{(2)}\equiv A_\mu$, a redefinition 
$k/(2\pi)\equiv \kappa$ and some partial integrations, it is straightforward to show that the action (\ref{nrabjm}) reduces to
\bea
S&=&+N(N-1)\int d^3x\left[-\frac{\kappa \hbar}{2}\epsilon^{\mu\nu\rho} A_\mu F_{\nu\rho}+\phi^*(i\hbar D_t)\phi-\frac{\hbar^2}{2m}|D\phi|^2+
\psi^* (i\hbar D_t)\psi\right.\cr
&&\left.-\frac{\hbar^2}{2m}|D\psi|^2+\frac{F_{12}}{2m}|\psi|^2-\frac{\hbar^2}{2\kappa m}|\phi|^2|\psi|^2-\frac{\hbar^2}{2m\kappa}|\phi|^4\right].
\eea
Further, noting that 
$
\epsilon^{\mu\nu\rho}A_\mu F_{\nu\rho}=\frac{2}{c}A_0 F_{12}-\epsilon^{ij}A_i \frac{1}{c}\d_0 A_j\,,
$
replacing our $(A_0)/c$ with $A_0$, and denoting $F_{12}=B$, we get precisely eq. (2.8) of \cite{Leblanc:1992wu}. The Yukawa 
term and scalar potential take the form
\be
\lambda_1|\phi|^2|\psi|^2 + \lambda_2|\phi|^4,
\ee
In particular, with $e\equiv 1$, we identify 
\be
  \lambda_1=\lambda_2=-\frac{\hbar^2}{2m\kappa}.
  \label{lambdaours}
\ee
Note that this combination of constants is {\it not} the one considered for the ${\cal N}=2$ supersymmetric case in  \cite{Leblanc:1992wu}, where rather
\be
  \lambda_1=+\frac{\hbar^2}{2mc\kappa};\;\;\;\;
  \lambda_2=3\lambda_1.\label{lambdatheirs}
\ee
However, in {\it both} cases, the Yukawa and self-interaction couplings satisfy the condition 
\be
2\lambda_1-\lambda_2+\frac{1}{2m\kappa}=0\;,
\ee
a necessary condition for ${\cal N}=1$ supersymmetry. In  \cite{Leblanc:1992wu}, it was further claimed that the condition
(\ref{lambdatheirs}) is the only solution to the ${\cal N}=2$ supersymmetry invariance. We disagree. In fact, we obtain the same 
supersymmetry tranformation laws, with the identification 
\be
\epsilon_1^{\rm ours}=-\sqrt{2m}\epsilon_1^{\rm theirs},\;\;\;\;\epsilon_2^{\rm ours}=i\sqrt{2m}\epsilon_2^{\rm theirs}\,,
\ee
and claim that they have simply not considered the case $\kappa<0$, which will result in our solution, as we now explain.

\noindent
In fact, there are {\it two} possible Bogomolnyi bounds, which arise from being able to write the Hamiltonian in two ways as a sum 
of complete squares plus a topological term,
\bea
\frac{1}{N(N-1)}{\cal H}&=&\frac{\hbar^2}{2m}\left[|D_\pm \phi|^2+|D_\pm \psi|^2\right]\pm \frac{\hbar^2}{2}\vec{\nabla}\times [\vec{j}_B+\vec{j}_F]
\pm \frac{\hbar^2}{4m}\vec{\nabla}^\rho_F\cr
&&-\left[\lambda_1\pm\frac{\hbar^2}{2m\kappa}\right]\rho_B^2-\left[\lambda_2\pm \frac{\hbar^3}{m\kappa}-\frac{\hbar^3}{2m\kappa}\right]\rho_B\rho_F\;,
\eea
with bosonic and fermionic currents
\bea
\vec{j}_B&=&\frac{1}{2mi}[\phi^*\vec{D}\phi-(\vec{D}\phi)^*\phi]\cr
\vec{j}_F&=&\frac{1}{2mi}[\psi^*\vec{D}\psi-(\vec{D}\psi)^*\psi+i\vec{\nabla}\times \rho_F].
\eea
If the fields are sufficiently well behaved, the integrals over the $j_B,j_F$ and $\rho_F$ terms vanish. If, in addition, the couplings
\be
  \lambda_1=\mp\frac{\hbar^2}{2m\kappa};\;\;\;
  \lambda_2=(1\mp 2)\frac{\hbar^3}{2m\kappa}\;,
\ee
the Hamiltonian reduces to 
\be
  H=\int d^2x\;\;
  \frac{\hbar^2}{2m}[|D_\pm \phi|^2+|D_\pm \psi|^2]\;,
\ee
which reaches its minimum value, zero, when the BPS equations
\bea
  D_1\phi&=&\mp i D_2\phi;\quad
  D_1\psi=\mp iD_2\psi.
  \label{BPS-system}
\eea
are satisfied. This choice of couplings clearly includes both our set, as well as that of \cite{Leblanc:1992wu}. Since, by the Olive-Witten 
theorem, these Bolgomolnyi equations are implied by the supersymmetry algebra in a supersymmetric theory, each BPS bound 
corresponds to specific set of supersymmetry transformations. This substantiates our claim above.\\

\noindent
The vortex solutions of the BPS system (\ref{BPS-system}) are easily extracted via the ansatz
\bea
  \phi&=&e^{i\theta_B}\rho_B^{1/2};\quad
  \psi=\eta\, e^{i\theta_F}\rho_F^{1/2}\;,
\eea
where $\eta$ is a constant spinor. As in the usual Jackiw-Pi case, these equations can be combined (using the fact that the 
fermionic and bosonic densities must be proportional) to
produce the Liouville equation
\be
  \nabla^2\ln\rho=\pm \frac{2}{\kappa}\rho.
\ee
This equation admits finite energy solutions only when the right hand side is negative as, for example when the lower sign is chosen 
with $\kappa>0$, as in \cite{Leblanc:1992wu}. However, and this is the subtlety that was not fully appreciated in \cite{Leblanc:1992wu}, 
it is also possible to have finite energy solutions by choosing the {\it upper sign} and $\kappa < 0$, as we have. At the level of the action, 
this corresponds to a parity transformation, which in turn leads to a supersymmetric theory with different couplings, BPS equations and 
solutions in a perfectly consistent way.

\subsection{Symmetries}
The symmetry algebra of our action, reduced to the supersymmetric Jackiw-Pi model is the same as in \cite{Leblanc:1992wu}, even with
the differing choice of couplings. Indeed, the algebra is independent of the values of the $\lambda_i$'s. For completeness, we review it here.\\

The reduced theory is invariant under the following bosonic symmetry operators: the Hamiltonian $H$, momentum $\vec{P}$, 
Galilean boost $\vec{G}$ , angular momentum $J_{12}$, dilation $D$, special conformal transformation $K$ and number operator $N$.
These symmetry operators satisfy the conformal Galilean algebra,
\bea
&&[P^i,P^j]=[P^i,H]=[J,H]=[G^i,G^j]=0,\nn\\
&& [J,P^i]=\epsilon^{ij} P^j\Rightarrow [M_{ij},P_k]=i(\delta_{ik} P_j-\delta_{jk}P_i),\nn\\
&& [J,G^i]=\epsilon^{ij} G^j\Rightarrow [M_{ij},G_k]=i(\delta_{ik} G_k-\delta_{jk}G_k),\nn\\
&& [P^i, G^j]=\delta^{ij} m N\equiv i\delta_{ij}\tilde N,     \nn\\
&&[H,G^i]=P,     \nn\\
&& [D,H]=-H,  \\   
&&[D,K]=K,         \nn\\
&&[H,K]=2D,       \nn\\
&&[K,J]=[K,G^i]=[D,J]=0,\nn\\
&&[K,P^i]=-G^i,\nn\\
&&[D,P^i]=-\frac{1}{2} P^i,\nn\\
&&[D,G^i]=\frac{1}{2} G^i.\nn
\eea
Here $i\tilde N\equiv mN$ is a mass operator that acts as a central charge. This is in excellent agreement with the conformal Galilean symmetry algebra considered in \cite{Balasubramanian:2008dm} (see also \cite{Son:2008ye}) for
$z=2$, i.e. the {\em Schr\"{o}dinger algebra}, with the identifications
\be
D=\frac{1}{2}i\tilde D;\;\;\;\;
K=-C;\;\;\;
M_{12}=iJ;\;\;\;
G^i=iK^i\;,
\ee
where $\tilde D, C, K^i, M_{ij}$ are the operators in \cite{Balasubramanian:2008dm}. The more general relations 
\be
[\tilde D,K_i]=(1-z)iK_i;\;\;\;
[\tilde D,H]=ziH
\ee
reduce to the above for $z=2$. Finally, the symmetry operators (in our notation, and for our action) are:
\bea
H&=&N(N-1)\int d^2x \left[\frac{\hbar^2}{2m}\left(|D\phi|^2+|\D\psi|^2\right)-\frac{B}{2m}\rho_F+\frac{\hbar^3}{2m\kappa}\rho_B\rho_F
+\frac{\hbar^2}{2m\kappa}\rho_B^2\right]\cr
P^i&=&\int d^2x\,{\cal P}^i=\frac{1}{2i}\int d^2x \left[\phi^*D^i\phi-(D^i\phi)^*\phi+\psi^*D^i\Psi-(D^i\psi)^*\psi\right]\cr
J&=&\int d^2 x\left[\vec{r}\times \vec{{\cal P}}+\frac{\rho_F}{2}\right]\cr
N&=&\int d^2x\,[\rho_B+\rho_F]\cr
G^i&=&t P^i-m\int d^2x\,[r^i(\rho_B+\rho_F)]\cr
D&=&t H-\frac{1}{2}\int d^2x\; \vec{r}\cdot \vec{\cal P}\cr
K&=&-t^2H+2tD+\frac{m}{2}\int d^2x\;\; r^2(\rho_B+\rho_F)\;,\label{symmop}
\eea
and where $\rho_B = |\phi|^2$ and $\rho_F = |\psi|^2$. We note also that the system is also invariant under two supersymmetries, 
which, together with the above form a supergroup, of ${\cal N}=2$ supersymmetric
Schr\"{o}dinger symmetry. Then the mass operator, which previously appeared only as a central charge, splits into bosonic and fermionic 
parts, $N_B$ and $N_F$, 
\be
N_B=\int d^2x\,\rho_B;\;\;\;
N_F=\int d^2x\,\rho_F\;,
\ee
which, together with a new generator $F$ coming from the commutator of the supercharge $Q_{2}$ with the generator of special conformal 
transformations, gives a total of 16 generators of the Super-Schr\"{o}dinger algebra. In fact, since only the explicit form of the generators 
$H, \vec{P}, J, \vec{G}, N_B, N_F, D$ and $K$ in terms of the fields are modified with respect to (\ref{symmop}) in the full action {\it before} 
truncation to the ${\cal N}=2$ Lagrangian of 
\cite{Leblanc:1992wu} and not their number, the symmetry algebra of the full theory with 4 complex scalars and 4 fermions is the same.

\section{AdS/CMT applications}

\subsection{Comments on systems with Super-Schr\"{o}dinger symmetry}

The appearance of this Super-Schr\"{o}dinger symmetry is remarkable in the context of the AdS/Condensed matter correspondence in two ways:
\begin{itemize}
\item
 It is, as far as we are aware, the first explicit example with an action, of a system with Schr\"{o}dinger (or in fact with any 
 conformal Galilean) algebra derived from a well-defined AdS/CFT duality in a ``top-down" way, {\it i.e.} embedded in a 
 critical string background (compare this, for example, to the nonlocal dipole theory constructed in \cite{Maldacena:2008wh}) and that 
 could be derived by reduction from 3+1 dimensions\footnote{We have not shown that explicitly, however note that the symmetry algebra
 can be embedded in 3+1 dimensions, and the theory is abelian.}, as is probably required for a good interpretation as a condensed matter 
 model.
\item
It is also a concrete example of a nonrelativistic AdS/CFT duality where the gravity dual is, as usual, $(d+1)-$dimensional and 
not $(d+2)-$dimensional. Indeed, the other concrete example of nonrelativistic AdS/CFT derived from a known duality was
constructed in \cite{Maldacena:2008wh,Herzog:2008wg,Adams:2008wt} by taking a discrete light cone quantization (DLCQ) 
of a known AdS/CFT pair, and in so doing killing one more coordinate (say, $x^+$, leaving an $x^-$), in addition to the radial coordinate $r$, 
leaving a duality between ``$CFT_d$" and ``$AdS_{d+2}$". For instance, the relevant case addressed in those works is the limit of the canonical 
$AdS_5\times S^5/{\cal N}=4$ SYM duality, leading to a duality between a 5-dimensional gravity dual and a 3-dimensional 
field theory with Schr\"{o}dinger symmetry. 
\end{itemize}
To summarise then, here we have a duality between a (2+1)-dimensional condensed matter system with a well-defined action and a 
certain limit of a 4-dimensional gravity dual, corresponding to massive ABJM (a deformation of the $AdS_4$ dual of the pure ABJM). 
While it is true that we still do not understand fully the effect of the abelian reduction nor of the norelativistic limit on the gravity dual; 
since the starting point was a conventional 4-dimensional gravity dual, we do anticipate that it will remain true of the endpoint 
as well\footnote{At this point, it is worth noting that a nonrelativistic limit of massive ABJM was 
taken, and a super-Schr\"{o}dinger symmetry was found in \cite{Lee:2009mm,Nakayama:2009cz}.  There too (see, for example, 
section 3.5 of \cite{Lee:2009mm}) it was noted 
that only an ${\cal N}=2$ subset (as is ours) could be embedded in a {\em four dimensional} relativistic superconformal symmetry,
via DLCQ (as in \cite{Maldacena:2008wh,Herzog:2008wg,Adams:2008wt}). Their full action and symmetry group cannot be 
embedded in 3+1 dimensions.}\footnote{Also note that gravitational spacetimes with the ${\cal N}=2$ Super-Schr\"{o}dinger 
symmetry have been found in \cite{Ooguri:2009cv,Jeong:2009aa}, however they still could only be interpreted as holographic duals 
in {\em two} dimensions lower.}.\\

\noindent
Finally, one may question how was it possible to obtain a system with Schr\"{o}dinger (conformal Galilean) symmetry, when we 
started from a system with a mass term (the non-conformal massive ABJM theory)? This is curious, but perfectly consistent even 
though after taking an abelian reduction and a nonrelativistic limit on the above and obtaining a theory with mass parameter 
$m$ since, in the nonrelativistic limit,  we can define units such that 
$\hbar $ {\it and} $m$ are dimensionless. In other words, $[t]=[r^2]$ and the dilatation symmetry is defined as 
\be
\delta t=2\a t;\;\;\;
\delta\vec{r}=\a \vec{r}.
\ee

\subsection{Comparison with nonrelativistic abelian toy models for ABJM}

In this penultimate section of the article, we will reflect on some more phenomenological aspects of the theory, keeping in mind our 
ultimate goal of building a concrete AdS/CMT correspondence embedded into a critical AdS/CFT duality. We will focus in particular 
on the physics of compressible quantum matter. In an interesting recent work \cite{Huijse:2011hp}, Huijse and Sachdev, initiated a 
study of compressible Fermi surfaces in as close to a ``top-down" approach as we have yet encountered. Their models were drawn 
from the canonical AdS/CFT duals ({\it viz} the 3-dimensional $\mathcal{N}=6$ ABJM and 4-dimensional $\mathcal{N}=4$ SYM theories) 
but even here, the paradigmic actions were taken only as a guide to developing a stable toy model. We would like to be able to do better.\\

\noindent
To that end, and for comparison, we write here the expression for action for the toy model proposed in  \cite{Huijse:2011hp},
\bea
S&=&\int d^3x\left[f_+^\dagger\left((\d_\tau-iA_\tau)-\frac{(\vec{\nabla}-i\vec{A})^2}{2m_f}-\mu\right)f_+\right.\cr
&&\left. +f_-^\dagger\left((\d_\tau+iA_\tau)-\frac{(\vec{\nabla}+i\vec{A})^2}{2m_f}-\mu\right)f_-\right.\cr
&&\left.+b_+^\dagger\left((\d_\tau-iA_\tau)-\frac{(\vec{\nabla}-i\vec{A})^2}{2m_b}+\epsilon_1-\mu\right)b_+\right.\cr
&&\left.+b_-^\dagger\left((\d_\tau+iA_\tau)-\frac{(\vec{\nabla}+i\vec{A})^2}{2m_b}+\epsilon_1-\mu\right)b_-\right.\cr
&&\left.+\frac{u}{2}(b_+^\dagger b_++b_-^\dagger b_-)^2+vb_+^\dagger b_-^\dagger b_-b_+ -g_1(b_+^\dagger b_-^\dagger f_-f_+ +h.c.)\right.\cr
&&\left.+c^\dagger\left(\d_\tau-\frac{(\vec{\nabla})^2}{2m_c}+\epsilon_2-\mu\right)c-g_2(c^\dagger(f_+b_-+f_-b_+)+h.c.)\right].
\label{hsmodel}
\eea
Here $A_\mu$ is an emergent gauge field ({\it i.e.}, not the electromagnetic gauge field), corresponding to a local $U(1)$ symmetry. Importantly, the theory also possesses a {\it global}U(1) symmetry with corresponding charge, 
\be
Q=f_+^\dagger f_++f_-^\dagger f_-+b_+^\dagger b_++b_-^\dagger b_-+2c^\dagger c,\label{hselcharge}
\ee
so both fundamental charged bosons $b_\pm$ and fermions $f_\pm$ as well as a neutral fermion $c$ all couple to the gauge field.

Since the kinetic terms for the fields are guaranteed to be the correct ones, we will instead focus on the scalar potential and Yukawa terms, whose sum we will denote by $V$. Before the nonrelativistic limit,
\bea
\frac{2}{N(N-1)}V=V_{mass}+V_{fer}+V_{quar}+V_{sext},
\eea
where (with $\tilde m \equiv mc/\hbar$, $\tilde k =k\hbar c$)
\bea
V_{mass} &=& \tilde m\sum_{i=1,2}\Big[\bar \eta_i\eta_i + \bar{\tilde \eta}_i\tilde\eta_i\Big] + \tilde m^2\Big[\left|\chi _1\right|^2 + \left|\chi _2\right|^2 
+ \left|\phi _1\right|^2 + \left|\phi _2\right|^2\Big],\\
V_{fer} &=& -\frac{2\pi i}{\tilde k}\Big[(|\phi_1|^2+|\chi_1|^2)(\bar\eta_2\eta_2+\bar{\tilde\eta}_2\tilde\eta_2)
+(|\phi_2|^2+|\chi_2|^2)(\bar{\eta}_1\eta_1+\bar{\tilde\eta}_1\tilde\eta_1)\Big],\\
V_{quar} &=& \frac{8 \pi \tilde m}{\tilde k}\Big[\left|\phi_1\right|^2\left|\phi_2\right|^2 - \left|\chi_1\right|^2\left|\chi_2\right|^2\Big],\\
V_{sext} &=& \frac{4\pi^2}{\tilde k^2}\Big[(\left|\chi_1\right|^2+\left|\phi_1\right|^2)(\left|\chi_2\right|^2+\left|\phi_2\right|^2) (\left|\chi_1\right|^2
+\left|\chi_2\right|^2+\left|\phi_1\right|^2+\left|\phi_2\right|^2)\Big].
\eea
The limit was defined by 
\bea
\Phi_b\longrightarrow\frac{\hbar}{\sqrt{2m}}\Phi_be^{-imc^2t/\hbar}\,\, ,\,\, \Psi_f\longrightarrow\sqrt{\hbar c}\Psi_fe^{-imc^2t/\hbar},
\eea
where $\Phi_b,\Psi_f$ are generic bosonic and fermionic fields, respectively. Of these, the mass terms are cancelled by the 
contributions of the mass in the exponent of the fields, the sextic term goes to zero, and for the rest we get
\bea
V^{NR}_{fer} &=& -\frac{\pi i\hbar^2}{mk}\Big[(|\phi_1|^2+|\chi_1|^2)(\eta^\dagger_2\eta_2+\tilde\eta^\dagger_2\tilde\eta_2)
+(|\phi_2|^2+|\chi_2|^2)(\eta^\dagger_1\eta_1+\tilde\eta^\dagger_1\tilde\eta_1)\Big],\nn\\
{}\\
V^{NR}_{quar} &=& \frac{2\pi\hbar^2}{k m}(\left|\phi_1\right|^2\left|\phi_2\right|^2-\left|\chi_1\right|^2\left|\chi_2\right|^2),\nn
\eea
after the nonrelativistic limit. Now, comparing this with the toy model (\ref{hsmodel}) we note that
\begin{enumerate}
 \item we are unable to obtain objects of all 3 charges (+,$-$ and 0) simultaneously. 
 \item we have $u=g_1=g_2=0$ and $\epsilon_1=\epsilon_2=\mu$ and $m_b=m_f=m$, and
 \item we obtain additional terms of the form $b^\dagger b f^\dagger f$. 
\end{enumerate}
With the truncation $A_\mu^{(1)}=-A_\mu^{(2)}=A_\mu$ and recalling that the covariant derivative $D_\mu = (\d_\mu-iA_\mu^{(i)})$ 
acting on both $\phi_{i}$ and $\psi_{i}$, we obtain the action
\bea
S&=&N(N-1)\int d^3x\left[\frac{k\hbar}{4\pi}\epsilon^{\mu\nu\rho}A_\mu F_{\nu\rho}+f_+^\dagger\left((\d_\tau-iA_\tau)-\frac{(\vec{\nabla}-i\vec{A})^2}{
2m}-\frac{F_{12}}{2m}\right)f_+\right.\nn\\
&&\left.+f_-^\dagger\left((\d_\tau+iA_\tau)-\frac{(\vec{\nabla}+i\vec{A})^2}{
2m}+\frac{F_{12}}{2m}\right)f_-
+b^\dagger_+\left((\d_\tau-iA_\tau)-\frac{(\vec{\nabla}-i\vec{A})^2}{2m}\right)b_+\right.\nn\\
&&+b^\dagger_-\left((\d_\tau+iA_\tau)-\frac{(\vec{\nabla}+i\vec{A})^2}{2m}\right)b_-
-\frac{2\pi \hbar^2}{mk}b^\dagger_+b^\dagger_-b_+b_-\nn\\
&&\left.-\frac{\pi \hbar^2}{km}[b^\dagger_+b_+f^\dagger_-f_-+b^\dagger_-b_-f^\dagger_+f_+]\right],
\eea
with an extra CS term, and where to facilitate comparison to (\ref{hsmodel}) we have denoted $\psi_1=f_+, \psi_2=f_-,\phi_1=b_+,\phi_2=b_-$ 
and changed to the conventions 
of \cite{Huijse:2011hp}. In other words, $v=-2\pi \hbar^2/(mk)$, $c=0$ and we also have some new couplings. Finally, it is worth noting that the matching 
is only consistent either for $f_-=0$, $\mu =F_{12}/(2m)$, or for $\mu=F_{12}=0$.\\ 

On the other hand, if we implemented the truncation by setting $A_\mu^{(2)}=0$, we would have no CS term, and $f_-=b_-=0$, producing the action, 
\footnote{Here, we set also $\psi_2=0$ and denoting $\psi_1=f_+,\phi_1=b_+,\phi_2=c$}
\bea
S&=&N(N-1)\int d^3x\left[f_+^\dagger\left((\d_\tau-iA_\tau)-\frac{(\vec{\nabla}-i\vec{A})^2}{
2m}-\frac{F_{12}}{2m}\right)f_++c^\dagger \left(\d_\tau-\frac{\vec{\nabla}^2}{2m}\right)c\right.\nn\\
&&\left.+b^\dagger_+\left((\d_\tau-iA_\tau)-\frac{(\vec{\nabla}-i\vec{A})^2}{2m}\right)b_+
-\frac{\pi \hbar^2}{km}c^\dagger c f^\dagger_+f_+\right.\nn\\
&&\left.-\frac{2\pi\hbar^2}{mk}b^\dagger_+b_+c^\dagger c\right].
\eea
Either way, the top-down model that we obtain does {\it not} match perfectly that of \cite{Huijse:2011hp}. Evidently then, while the 
mathematical structure of the two models are strikingly similar, their differences are sufficient to warrant further development, and 
it remains to be seen how much of the condensed matter physics can actually be reproduced. 

\section{Conclusions}

As part of a more ambitious progam aimed at a full {\it top-down} realization of the AdS/CMT correspondence in the AdS/CFT duality, this article 
details our analysis of the nonrelativistic limit of the abelian reduction of the massive ABJM model proposed in \cite{Murugan:2013jm}. We have also 
checked that this limit commutes with our abelianization procedure (though we did not present the details here). 
Moreover, in our study of the supersymmetry laws governing 
the nonrelativistic limit, we found that the scaling of the supersymmetry parameters is not unique. Either we can keep all the fields, and the 
supersymmetry laws become rather simple and involve only a kinematical piece. However, the price we pay is that the resulting scalar 
potential is unbounded from below. Alternatively, we can truncate the theory to 2 complex fermions and 2 complex scalars, and obtain a system with 
${\cal N}=2$ supersymmetry with both kinematical and dynamical susy pieces. 
Further truncation to a single complex fermion and a single complex scalar yields a 
supersymmetric version of the Jackiw-Pi model considered in \cite{Leblanc:1992wu}, although with novel values for the parameters.\\

\noindent
The system we obtain has Super-Schr\"{o}dinger symmetry and constitutes a concrete example of an interesting condensed matter model 
with an explicit action, obtained as a limit of a known AdS/CFT duality. Moreover, the holographic duality here is of the conventional type 
related to a ($d+1$)-dimensional gravity theory, instead of the previously constructed $(d+2)-$dimensional holographic dual.\\ 

\noindent
Finally, on a more phenomenological note, we have compared our top-down construction with previously used abelian nonrelativistic condensed 
matter avatars for the ABJM model \cite{Huijse:2011hp} that were explored in the context of compressible quantum matter, 
and found that there are certainly differences, the similarities between the models is striking. Indeed, it would be intriguing to push these 
similarities and see just how much of the physics of quantum matter can be teased from the nonrelativistic abelianized ABJM model. 
We leave this as an invitation to future work.

\section{Acknowledgements}

We would like to thank Asadig Mohammed for collaboration on early stages of this work and all the participants, 
but especially Johanna Erdmenger, Aki Hashimoto and Koenraad Schalm,
of the ``Frontiers of the AdS/CMT correspondence" workshop in Cape Town, 2014 where this work was initiated. 
The work of HN is supported in part by CNPq grant 301219/2010-9 and FAPESP grants 2013/14152-7 and 2014/18634-9. 
JM and CLA acknowledge support from the National Research Foundation (NRF) of South Africa's CPRR program under GUN 87667.

\bibliography{nrabjm}
\bibliographystyle{utphys}

\end{document}